\documentclass[epjc3]{svjour3} 

\usepackage[margin=25mm]{geometry}
\usepackage{amsmath,amssymb}

\usepackage[dvipdf]{graphicx}
\usepackage{subfigure}
\usepackage{ulem}
\usepackage{cite}
\usepackage{url} 
\usepackage{color}

\newcommand{\Peq}{P_{\mathrm{eq}}}
\newcommand{\Teq}{T_{\mathrm{eq}}}

\newcommand{\ST}{S_{q}}
\newcommand{\SR}{S_{\mathrm{R}q}}

\newcommand{\hH}{\hat{H}}
\newcommand{\hA}{\hat{A}}
\newcommand{\hrho}{\hat{\rho}} 
\newcommand{\Tr}{\mathrm{Tr}}

\newcommand{\Corr}{\mathrm{Corr}}
\newcommand{\fsc}{f^{\mathrm{sc}}(p_N^{\mathrm{sc}})}
\newcommand{\PRi}{P_{\mathrm{R}i}}
\newcommand{\PRj}{P_{\mathrm{R}j}}
\newcommand{\aR}{\alpha_{\mathrm{R}}}
\newcommand{\bR}{\beta_{\mathrm{R}}}

\title{Momentum distribution and correlation function of free particles in the Tsallis statistics using conventional expectation value and equilibrium temperature}
\author{Masamichi Ishihara\thanksref{e1,addr1}}
\thankstext{e1}{email: masamichi.ishihara.research@gmail.com}
\institute{Department of Economics, Faculty of Economics, Chiba Keizai University, Chiba, 263-0021, Japan \label{addr1}}

\begin{document}
\maketitle
\begin{abstract}
  We applied the Tsallis statistics with the conventional expectation value to a system of free particles,  
  adopting the equilibrium temperature, which is often called the physical temperature.
  The entropic parameter $q$ in the Tsallis statistics is less than one for power-law-like distribution.
  The well-known relation between the energy and the temperature in the Boltzmann--Gibbs statistics holds in the Tsallis statistics,  
  when the equilibrium temperature is adopted. 
  We derived the momentum distribution and the correlation function in the Tsallis statistics. 
  The momentum distribution and the correlation function in the Tsallis statistics are different from those in the Boltzmann--Gibbs statistics,
  even when the equilibrium temperature is adopted.
  These quantities depend on $q$ and $N$, where $N$ is the number of particles. 
  It is also shown that the parameter $q$ is required to satisfy $1-1/(3N/2+1) < q < 1$. 
\end{abstract}

\section{Introduction}

Power-law-like distributions appear in various branches of science.
The Tsallis statistics \cite{Tsallis:1988, Tsallis:Book:2010} is a candidate to describe the distributions,
such as momentum distributions of particles \cite{Parvan:PLA:2006, Bhattacharyya:Tsallis-1:2025}.
The Tsallis statistics is based on the Tsallis entropy and the expectation value.
There are three types of the expectation values \cite{Tsallis:PhysicaA:1998}:
the conventional expectation value, the $q$-expectation value,
and the normalized $q$-expectation value which is often called the escort average.

The probability or the density operator is obtained under the maximum entropy principle \cite{Jaynes:1957}.
The Tsallis entropy is extremized with both the normalization condition and the energy constraint.
The entropic parameter $q$ for a power-law-like distribution is less than one in the Tsallis statistics
with the conventional expectation value.
The probability or the density operator in the Boltzmann--Gibbs statistics is recovered
when $q$ approaches one from below. 

The equilibrium temperature
\cite{Ishihara:2024:EPJP139,Imdiker:EPJC:2023,Ishihara:EPJP:2023:1,Ishihara:EPJP:2023:2,Ishihara:arXiv:2025},
which is often called the physical temperature
\cite{Abe-PLA:2001,Kalyana:2000,S.Abe:physicaA:2001,Aragao:2003,Ruthotto:2003,Toral:2003,Suyari:2006,Ishihara:phi4,Ishihara:free-field,Ishihara:EPJB:2022,Ishihara:EPJB:2023,Ignatyuk_2024},
is introduced in the Tsallis statistics.
The inverse of the conventional temperature is defined by the partial derivative of the entropy with respect to the energy, 
and is related to the Lagrange multiplier of the functional under the maximum entropy principle.
The equilibrium temperature is introduced similarly by referring the property of the Tsallis entropy.
This temperature is related to the R\'enyi entropy, which is expressed in terms of the Tsallis entropy.

The Tsallis statistics, which adopts the notion of the equilibrium temperature and incorporates the conventional expectation value, 
may be preferable for describing power-law-like distributions.
The expectation value of the unit operator $\hat{1}$ is one
in the Tsallis statistics with the conventional expectation value and
in the Tsallis statistics with the normalized $q$-expectation value. 
The equilibrium temperature characterizes the system.
The description using the equilibrium temperature is valid within the framework of the Tsallis statistics.

The purpose of this paper is to obtain the momentum distribution and the correlation function
in a system of free particles in the Tsallis statistics.
The conventional expectation value is adopted and the quantities are expressed in terms of the equilibrium temperature. 
The well-known relation between the energy and the temperature is shown.
It is shown that the momentum distribution is power-law-like even when the equilibrium temperature is adopted.
The correlation function depends on the magnitudes of the momenta. 
It is also shown that the parameter $q$ is required to satisfy $1-1/(3N/2+1)<q<1$.
This condition is similar to the conditions previously reported
\cite{Abe-PLA:2001, Lenzi:PLA:2001, Parvan:PLA:2006, Parvan:PLA:2010, Ishihara:EPJB:2022,Ishihara:EPJB:2023,Ishihara:2024:EPJP139}.

This paper is organized as follows.
In Sect.~\ref{sec:Tsallis-1:prob:eq-tempe},
the Tsallis statistics with the conventional expectation value is briefly reviewed.
The equilibrium temperature, which is often called the physical temperature, is introduced.
In Sect.~\ref{sec:N_particle_sys}, the Tsallis statistics is applied to the $N$ free particle system.
The relation between the energy and the equilibrium temperature is derived. 
Moreover, the momentum distribution and the correlation function are calculated.
The last section is assigned for discussion and conclusions. 
In \ref{sec:appendix:Renyi}, the probabilty is given in the R\'enyi statistics.
In \ref{sec:appendix:Integrals}, the results of the integrals appearing in this paper are presented.

\section{The Tsallis statistics using the equilibrium temperature}
\label{sec:Tsallis-1:prob:eq-tempe}
We briefly review the Tsallis statistics. 
The probability or the density operator is obtained in the Tsallis statistics with the conventional expectation value:
the framework is often called the Tsallis-1 statistics \cite{Parvan:PLA:2006, Ishihara:2024:EPJP139, Ishihara:arXiv:2025, Parvan:PhysicaA:2022}.
The entropy and the expectation value have essential roles in the Tsallis statistics.
The Tsallis entropy $\ST$ in quantum statistics is given by
\begin{align}
  \ST = \frac{ \displaystyle \Tr(\hrho^q) - 1 }{1-q},
  \label{def:S:Tsallis}
\end{align}
where $\hrho$ is the density operator.
We employ the conventional expectation value. The expectation value of a quantity $\hA$ is given by
\begin{align}
\langle \hA \rangle = \Tr(\hA \hrho). 
\end{align}
The density operator $\hrho$ is obtained by applying the maximum entropy principle.
The functional with both the normalization condition and the energy constraint is defined by
\begin{align}
  I = \ST - \alpha \left[ \Tr\left( \hrho \right)  - 1 \right] - \beta \left[ \Tr(\hH\hrho) - U \right],
  \label{eqn:functional}
\end{align}
where $\hH$ is the Hamiltonian and $U$ is the energy.
The parameters, $\alpha$ and $\beta$, are the Lagrange multipliers.  
The density operator is obtained by requiring $\delta I = 0$ \cite{Ishihara:arXiv:2025}:
\begin{align}
\hrho = \left[ 1 + (1-q) \ST + \left( \frac{1-q}{q} \right) \beta (\hH - U) \right]^{1/(q-1)} . 
\end{align}
In the same way,  
we can obtain the probability $P_i$ for the Tsallis entropy $\ST = (\sum_i P_i^q - 1)/(1-q)$, applying the maximum entropy principle: 
\begin{align}
P_i = \left[ 1 + (1-q) \ST + \left( \frac{1-q}{q} \right) \beta (E_i - U) \right]^{1/(q-1)} , 
\end{align}
where $E_i$ is the energy in state $i$. 

We introduce the equilibrium temperature, which is often called the physical temperature. 
We represent the Tsallis entropy of subsystem A as $\ST^{\mathrm{A}}$, 
that of subsystem B as $\ST^{\mathrm{B}}$,
and that of the system composed of A and B as $\ST^{\mathrm{A+B}}$.
The Tsallis entropy satisfies the following relation for independent subsystems: 
\begin{align}
\ST^{\mathrm{A+B}} = \ST^{\mathrm{A}} + \ST^{\mathrm{B}} + (1-q) \ST^{\mathrm{A}} \ST^{\mathrm{B}} . 
\end{align}
We assume the additivity of the energy $U^{A+B} = U^{A} + U^{B}$. 
The requirements, $\delta \ST^{\mathrm{A+B}} =0$ and $\delta U^{\mathrm{A+B}} = 0$, gives
\begin{align}
  \frac{1}{1+(1-q) \ST^{\mathrm{A}}} \frac{\partial \ST^{\mathrm{A}}}{\partial U^{\mathrm{A}}}
  = \frac{1}{1+(1-q) \ST^{\mathrm{B}}} \frac{\partial \ST^{\mathrm{B}}}{\partial U^{\mathrm{B}}} .
\end{align}
The equilibrium temperature $\Teq$ is introduced by 
\begin{align}
  \frac{1}{\Teq} = \frac{1}{1+(1-q) \ST} \frac{\partial \ST}{\partial U}.
\end{align}
The Lagrange multiplier $\beta$ is related to the derivative ${\partial \ST}/{\partial U}$
through the relation $\beta = {\partial \ST}/{\partial U}$.
Therefore, we have
\begin{align}
  \frac{1}{\Teq} = \frac{\beta}{R},
  \label{eqn:Teq:beta:R}
\end{align}
where $R$ is defined by $R = 1 +(1-q) \ST$.

The probability $P_i$ is expressed in terms of $R$ and $\Teq$:
\begin{align}
P_i = R^{ \frac{1}{q-1} } \left[ 1 + \left(\frac{1-q}{q} \right) \left( \frac{E_i - U}{\Teq} \right) \right]^{\frac{1}{q-1}} . 
\end{align}
The probability is rewritten without $R$ by using the relation $R = \sum_i (P_i)^q$.
The probability $P_i$ is given by \cite{Ishihara:arXiv:2025}
\begin{align}
  P_i = \frac{\left[ 1 + \left(\frac{1-q}{q} \right) \left(\frac{E_i - U}{\Teq}\right) \right]^{\frac{1}{q-1}} }
  {\displaystyle\sum_j \left[1 + \left(\frac{1-q}{q}\right) \left(\frac{E_j - U}{\Teq}\right) \right]^{\frac{q}{q-1}} }.
  \label{p_i}
\end{align}
The energy $U$ is contained in the probability $P_i$. 
The following condition should be satisfied for $0 < q < 1$: 
\begin{align}
1 + \left(\frac{1-q}{q}\right) \left(\frac{E_{\mathrm{min}} - U}{\Teq}\right) > 0 ,  
\label{condition:p}
\end{align}
where $E_{\mathrm{min}}$ is the lowest energy.
It may be worthwhile to mention that the probability $P_i$, Eq.~\eqref{p_i}, is also obtained in the R\'enyi statistics. 
The derivation is given in \ref{sec:appendix:Renyi}.

\section{The system of $N$ free particles under the Tsallis statistics}
\label{sec:N_particle_sys}
\subsection{The probability and the relation between the energy and the equilibrium temperature}
\label{subsec:free-particle:probability}
We treat $N$ free particles of mass $m$ in a cubic box of side length $L$.
The energy is
\begin{align}
E = \sum_{j=1}^{3N} \frac{p_j^2}{2m} . 
\label{Hamiltonian}
\end{align}
The Dirichlet boundary condition is imposed, and the momentum $p_j$ is given by 
\begin{align}
p_j = \left( \frac{\hbar\pi}{L} \right) n_j, 
\label{Quantization}
\end{align}
where $n_j$ is a positive integer.  
We introduce the variable $\varepsilon$ by
\begin{align}
\varepsilon = \frac{\hbar^2 \pi^2}{2mL^2} .
\end{align}
The energy $E$ can be expressed in terms of $\varepsilon$ as
\begin{align}
E = \varepsilon \sum_{j=1}^{3N} n_j^2 .
\label{eqn:energy_reexpressed}
\end{align}  

The index $i$ of the probability $P_i$ determines the set $(n_1, n_2, \cdots, n_{3N})$. 
Therefore, we can represent the probability $P_i$ as $P(n_1, n_2, \cdots, n_{3N})$.
The factor $1/N!$ associated with particle indistinguishability is absorbed into this probability.
We first examine the denominator $\rho_D$ of the probability which is given by
\begin{align}
  \rho_D = \frac{1}{N!} \left[1-\left(\frac{1-q}{q}\right)\frac{U}{\Teq}\right]^{\frac{q}{q-1}}
  \sum_{n_1, n_2, \cdots, n_{3N}=1}^{\infty}
  \left[
    1 +  \frac{\left(\frac{1-q}{q}\right) \frac{\varepsilon}{\Teq}}
    {\left(1-\left(\frac{1-q}{q}\right)\frac{U}{\Teq}\right)}
    (n_1^2 + n_2^2 + \cdots + n_{3N}^2)
    \right]^{\frac{q}{q-1}} .  
\end{align}
The lowest energy $E_{\mathrm{min}}$ goes to zero, as $L$ goes to infinity.
In such a case, we have $1-((1-q)/q)(U/\Teq) > 0$ from Eq.~\eqref{condition:p}.
Therefore, we assume that $1-((1-q)/q)(U/\Teq)$ is larger than zero.
This assumption is justified for large $L$.
We introduce a variable $x_j$ by
\begin{align}
  x_j = \left[ \frac{\left(\frac{1-q}{q}\right) \frac{\varepsilon}{\Teq}}{1-\left(\frac{1-q}{q}\right)\frac{U}{\Teq}} \right]^{1/2} n_j . 
\end{align}
Therefore, for large $L$,
$\rho_D$ is approximated as follows:
\begin{align}
  \rho_D = \frac{1}{N!} 
  \frac{\left[1-\left(\frac{1-q}{q}\right)\frac{U}{\Teq}\right]^{\frac{q}{q-1}+\frac{3N}{2}}}
       {\left[ \left(\frac{1-q}{q}\right) \frac{\varepsilon}{\Teq} \right]^{\frac{3N}{2}}}
       \int_0^{\infty} dx_1 \cdots dx_{3N} (1+x_1^2+\cdots+x_{3N}^2)^{\frac{q}{q-1}} .
\end{align}
The denominator $\rho_D$ is calculated:
\begin{align}
  \rho_D = \frac{1}{N!} \frac{\pi^{\frac{3N}{2}}}{2^{3N}}
  \frac{\left[1-\left(\frac{1-q}{q}\right)\frac{U}{\Teq}\right]^{\frac{q}{q-1}+\frac{3N}{2}}}
       {\left[ \left(\frac{1-q}{q}\right) \frac{\varepsilon}{\Teq} \right]^{\frac{3N}{2}}}
  \frac{\Gamma\left(\frac{q}{1-q} - \frac{3N}{2}\right) }{\Gamma\left(\frac{q}{1-q}\right)}, 
\end{align}
where $\Gamma(x)$ is the Gamma function. 
The following inequalities must be satisfied for the previous calculation to be valid: 
\begin{subequations}
\begin{align}
  & 1-\frac{1}{(3N/2+1)} < q < 1, \label{eqn:cond_denom_prob:1} \\
  & N>0. \label{eqn:cond_denom_prob:2}
\end{align}  
\label{eqn:cond_denom_prob}
\end{subequations}
Similar conditions can be obtained from the numerator of the probability $P(n_1, n_2, \cdots, n_{3N})$
because the normalization condition must be satisfied: 
\begin{align}
\sum_{n_1=1, \cdots, n_{3N}=1}^{\infty} P(n_1, \cdots, n_{3N})  = 1. 
\label{eqn:cond_nume_prob}
\end{align}
The conditions required for Eq.~\eqref{eqn:cond_nume_prob} to hold are satisfied
under Eqs.~\eqref{eqn:cond_denom_prob:1} and \eqref{eqn:cond_denom_prob:2}. 
Under these conditions, the probability $P(n_1,\cdots,n_{3N})$ for the present system can be written explicitly as
\begin{align}
  P(n_1,\cdots,n_{3N}) = \frac{1}{\rho_D}
  \frac{1}{N!} 
  &
  \left[
   \left(1-\left(\frac{1-q}{q}\right)\frac{U}{\Teq}\right)
    +
    \left(\frac{1-q}{q}\right) \frac{\varepsilon (n_1^2 + n_2^2 + \cdots + n_{3N}^2)}{\Teq}
    \right]^{\frac{1}{q-1}}.
  \label{eqn:Pn}
\end{align}

The relation between $U$ and $\Teq$ can be shown, and the Tsallis entropy can be also shown.
The energy $U$ is obtained using Eqs.~\eqref{eqn:energy_reexpressed} and \eqref{eqn:Pn}:
\begin{align}
  U  = \sum_{n_1, \cdots, n_{3N}=1}^{\infty}   \varepsilon (n_1^2+\cdots+n_{3N}^2) P(n_1, n_2, \cdots, n_{3N}) .
  \label{eqn:U}
\end{align}
The right-hand side of Eq.~\eqref{eqn:U} is calculated in the same way.
We have
\begin{align}
U =\frac{3}{2} N \Teq. 
\label{eqn:U-Teq}
\end{align}
This relation seems to be natural.
It may be worth mentioning that the energy is generally obtained by solving a self-consistent equation.
The Tsallis entropy $\ST$ is expressed in terms of $\rho_D$:
\begin{align}
  \ST = \frac{\rho_D^{1-q} -1}{1-q}.
  \label{eqn:ST_denom}
\end{align}
The expression for the Tsallis entropy in the limit $q \rightarrow 1$
coincides with the Sackur-Tetrode equation, which gives the Boltzmann--Gibbs entropy for an ideal gas. 
Using Eqs.~\eqref{eqn:Teq:beta:R} and \eqref{eqn:ST_denom}, we have
\begin{align}
\Teq = (1+(1-q)\ST) \beta^{-1} = \rho_D^{1-q} \beta^{-1}. 
\label{eqn:Teq:Sq:rhoD}
\end{align}
The conditions on $q$ and $N$ for $\rho_D$ guarantee the finiteness of the entropy and that of the equilibrium temperature. 
The quantity $\sum_i P_i^q$ is also finite because $\sum_i P_i^q$ is equal to $(1+(1-q)\ST)$.

\subsection{Momentum distribution in the system of $N$ free particles}
We focus on the momentum distribution $f^{[1]}(\vec{p}_N)$, which is the probability distribution of the momentum $\vec{p}_N$, 
in the Tsallis statistics with the conventional expectation value.
The quantity $\vec{p}_N$ is the momentum of the particle $N$.
We adopt the equilibrium temperature. 
We calculate the sum over $n_1, \cdots, n_{3(N-1)}$ of the probability.
The remaining variables are $n_{3N-2}, n_{3N-1}, n_{3N}$ which are related to the momentum $\vec{p}_N$.
We continue to consider the case of sufficiently large $L$.

We calculate the quantity $\sum_{n_{1}, \cdots, n_{3(N-1)} = 1}^{\infty} P(n_1, \cdots, n_{3N})$
to obtain the distribution $f^{[1]}(\vec{p}_N)$:
\begin{align}
  & \sum_{n_{1}, \cdots, n_{3(N-1)} = 1}^{\infty} P(n_1, \cdots, n_{3N})  
  \nonumber \\ &
  = \frac{2^3}{\pi^{3/2}} 
    \frac{ \left[\left(\frac{1-q}{q}\right) \frac{\varepsilon}{\Teq} \right]^{\frac{3}{2}}}
         {\left[ 1-\left(\frac{1-q}{q}\right) \frac{U}{\Teq} \right]^{\frac{5}{2}}}
    \frac{\Gamma\left(\frac{1}{1-q} - \frac{3}{2}(N-1)\right) }{\Gamma\left(\frac{q}{1-q} - \frac{3}{2} N \right)}
    \frac{\Gamma\left(\frac{q}{1-q}\right)}{\Gamma\left(\frac{1}{1-q}\right)}
    \left\{
      1 + \frac{\left(\frac{1-q}{q}\right)}{\left[1-\frac{1-q}{q}\frac{U}{\Teq}\right]} \frac{(\vec{p}_N)^2}{2m\Teq}
      \right\}^{\frac{1}{q-1}+\frac{3}{2}(N-1)},
\label{sum_of_P:momentum:positive}
\end{align}
where $\vec{p}_N = (p_{3N-2}, p_{3N-1}, p_{3N})$. 
The variables $p_{3N-2}$, $p_{3N-1}$, and $p_{3N}$ are positive in Eq.~\eqref{sum_of_P:momentum:positive},
because $n_{3N-2}$, $n_{3N-1}$, and $n_{3N}$ are positive. 
Certain conditions must be satisfied for the previous calculation to be valid. 
One is the condition $N>1$. The others hold when Eq.~\eqref{eqn:cond_denom_prob:1} is satisfied.

We require that the distribution $f^{[1]}(\vec{p}_N)$ satisfies
\begin{align}
\int_{-\infty}^{\infty} dp_{3N-2} dp_{3N-1} dp_{3N} f^{[1]}(\vec{p}_N) = 1, 
\end{align}
considering the relation $  \sum_{n_{3N-2}, n_{3N-1}, n_{3N} = 1}^{\infty} \left( \sum_{n_{1}, \cdots, n_{3(N-1)} = 1}^{\infty} P(n_1, \cdots, n_{3N}) \right) = 1$.
The sum over positive integers can be symmetrically extended to all integers
in the case of sufficiently large $L$, because $P(n_1,n_2, \cdots, n_{3N})$ is even in each argument.
The distribution $f^{[1]}(\vec{p}_N)$ should be given by 
\begin{align}
  f^{[1]}(\vec{p}_N)
  &= \frac{1}{2^3} \left(\frac{L}{\hbar\pi}\right)^3  \left( \sum_{n_{1}, \cdots, n_{3(N-1)} = 1}^{\infty} P(n_1, \cdots, n_{3N}) \right).
\end{align}
We obtain
\begin{align}
  f^{[1]}(\vec{p}_N) = 
    \frac{\left(\frac{1-q}{q}\right)^{\frac{5}{2}}}{\left[ 1-\left(\frac{1-q}{q}\right) \frac{U}{\Teq} \right]^{\frac{5}{2}}}
    \frac{\Gamma\left(\frac{1}{1-q} - \frac{3(N-1)}{2}\right) }{\Gamma\left(\frac{q}{1-q} - \frac{3N}{2} \right)}
    \frac{1}{(2\pi m\Teq)^{\frac{3}{2}}}
    \left[
      1 + \frac{\left(\frac{1-q}{q}\right)}{\left[1-\frac{1-q}{q}\frac{U}{\Teq}\right]} \frac{(\vec{p}_N)^2}{2m\Teq}
      \right]^{\frac{1}{q-1}+\frac{3(N-1)}{2}} ,
\label{eqn:momdist:1}
\end{align}
where $p_j$ for $j=3N-2, 3N-1, 3N$ satisfies $-\infty < p_{j} < \infty$ in Eq.~\eqref{eqn:momdist:1}.
Using the equation $U=3N\Teq/2$, we can rewrite the distribution as
\begin{align}
  f^{[1]}(\vec{p}_N) = 
    \frac{1}{\left[ \left(\frac{q}{1-q}\right)- \frac{3N}{2} \right]^{\frac{5}{2}}}
    \frac{\Gamma\left(\frac{1}{1-q} - \frac{3(N-1)}{2}\right) }{\Gamma\left(\frac{q}{1-q} - \frac{3N}{2} \right)}
    \frac{1}{(2\pi m\Teq)^{\frac{3}{2}}}
    \left[
      1 + \frac{1}{\left[\frac{q}{1-q}-\frac{3N}{2}\right]} \frac{(\vec{p}_N)^2}{2m\Teq}
      \right]^{\frac{1}{q-1}+\frac{3(N-1)}{2}} .
\label{eqn:momdist:2}
\end{align}
The distribution $f^{[1]}(\vec{p}_N)$ depends only on the magnitude $p_N = |\vec{p}_N|$ of the vector $\vec{p}_N$.
The distribution $f(p_N)$  which is defined by $4\pi (p_N)^2 f^{[1]}(\vec{p}_N)$ is
obviously given by using Eq.~\eqref{eqn:momdist:2}.
The distribution $f(p_N)$ approaches the well-known distribution in the Boltzmann--Gibbs statistics,
as $q$ approaches one from below: 
\begin{align}
  \lim_{q \rightarrow 1^{-}} f(p_N) \equiv \lim_{q \rightarrow 1^{-}} 4 \pi (p_N)^2 f^{[1]}(\vec{p}_N)
  = \frac{4\pi (p_N)^2}{(2\pi m\Teq)^{3/2}} \exp\left(-\frac{(p_N)^2}{2m\Teq}\right) .
\end{align}

The pressure $\Peq$ is calculated using $f^{[1]}(\vec{p}_N)$.
In the present case, we obtain
\begin{align}
\Peq = L^{-3} N \Teq . 
\end{align}
Using Eq.~\eqref{eqn:U-Teq}, we have
\begin{align}
\Peq V = \frac{2}{3} U, 
\end{align}
where $V$ is the volume of the box of side length $L$. 

For numerical illustration, 
we introduce the following scaled momentum:
\begin{subequations}
\begin{align}
  &\vec{p}_j^{\mathrm{sc}} = \vec{p}_j / \sqrt{m \Teq}, \\ 
  &p_j^{\mathrm{sc}} = |\vec{p}_j^{\mathrm{sc}}| . 
\end{align}
\label{eqn:scaled_varible}
\end{subequations}
We also introduce the following scaled distribution:
\begin{align}
  f^{\mathrm{sc}}(p_N^{\mathrm{sc}}) = \sqrt{m \Teq} f(p_N).
\end{align}

\begin{figure}
\centering
\subfigure[Momentum distributions for $q=0.97$, $0.985$, and $1.0$.
  The function at $q=1$ corresponds to the distribution in the Boltzmann-Gibbs statistics.]
          {
            \includegraphics[width=0.45\textwidth]{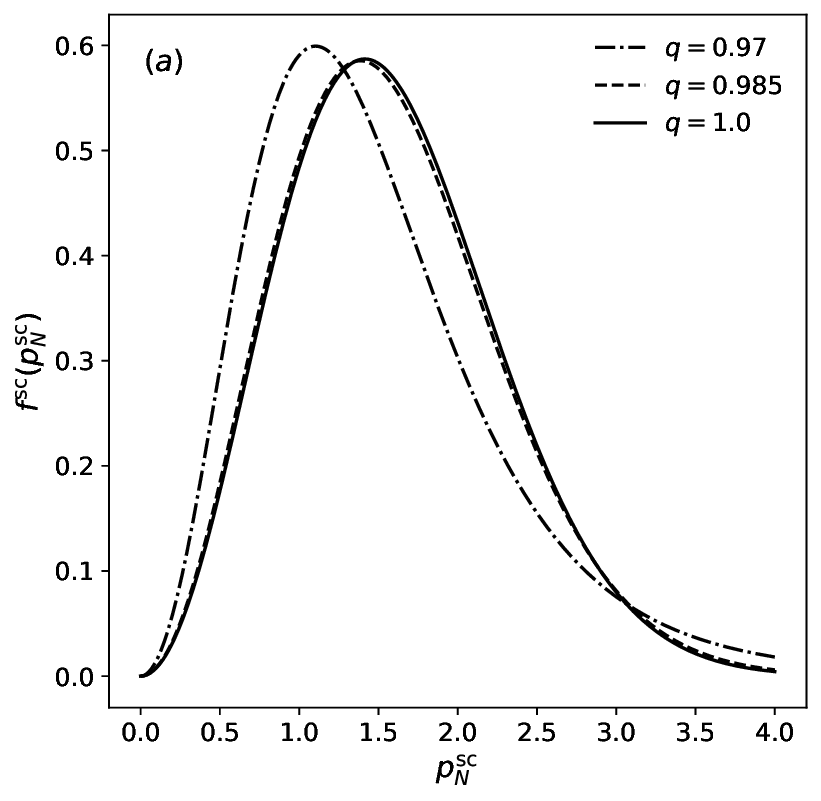}
            \label{fig:mom:wide}
          }
\quad
\subfigure[Momentum distributions for $q=0.968$, $q=0.97$, and $0.972$.]
          {
            \includegraphics[width=0.45\textwidth]{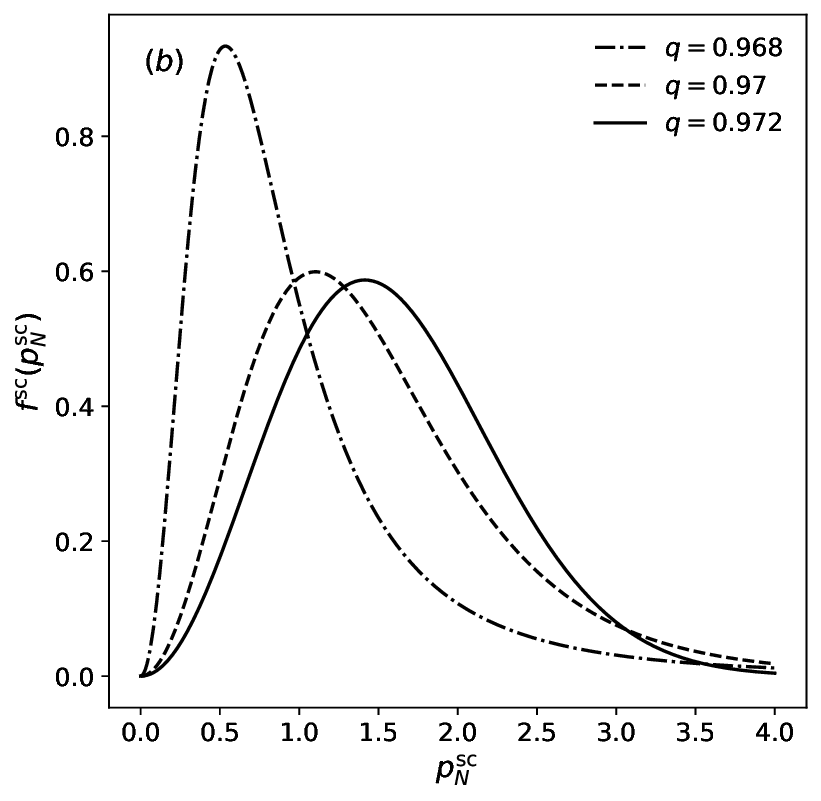}
            \label{fig:mom:narrow}
          }
\caption{Momentum distributions $f^{\mathrm{sc}}(p_N^{\mathrm{sc}})$ as functions of scaled momentum $p_N^{\mathrm{sc}}$ for various values of $q$ at $N=20$}
\label{fig:mom}
\end{figure}

Figures~\ref{fig:mom:wide}\ref{fig:mom:narrow} show the momentum distribution $\fsc$ for various values of $q$ at $N=20$.
In Fig.~\ref{fig:mom:wide}, 
the distribution at $q=0.985$ is slightly different from the distribution in the Boltzmann-Gibbs statistics ($q=1$).
It is easily found that the distribution at $q=0.97$ is different from that at $q=1$.
In Fig.~\ref{fig:mom:narrow}, 
the peak height increases with decreasing $q$, although the difference between $q=0.972$ and $q=0.97$ is relatively small. 
The distribution for $q=0.968$ is noticeably narrower than those for  $q=0.972$ and $q=0.97$.
The tail of the distribution can be seen at $q=0.968$. 
The lower bound of $q$ is obtained from Eq.~\eqref{eqn:cond_denom_prob:1}:
the boundary value of $q$ at $N=20$ is $30/31$.

\subsection{Correlation function in the system of $N$ free particles}
The distribution $f^{[2]}(\vec{p}_{N-1}, \vec{p}_{N})$,
which is the probability distribution for the momenta $\vec{p}_{N-1}$ and $\vec{p}_{N}$,
is calculated in the similar way, where $\vec{p}_{N-1} = (p_{3N-5}, p_{3N-4}, p_{3N-3})$ and $\vec{p}_{N} = (p_{3N-2}, p_{3N-1}, p_{3N})$. 
We continue to consider the case of sufficiently large $L$.
We have
\begin{align}
  \sum_{n_1, \cdots, n_{3(N-2)}=1}^{\infty} P(n_1, \cdots, n_{3N})
  =& \frac{2^6}{\pi^3} \left(\frac{1-q}{q}\right)
  \frac{\left[\frac{1-q}{q} \frac{\varepsilon}{\Teq}\right]^3}{\left[1-\frac{1-q}{q} \frac{U}{\Teq}\right]^4}
  \frac{\Gamma \left(\frac{1}{1-q}-\frac{3(N-2)}{2} \right)}{\Gamma\left(\frac{q}{1-q}-\frac{3N}{2}\right)}
  \nonumber\\ &  \times
  \left[
  1+
  \frac{\left(\frac{1-q}{q}\right)}{1-\left(\frac{1-q}{q}\right) \frac{U}{\Teq}}
  \frac{(\vec{p}_{N-1})^2+(\vec{p}_{N})^2}{2m\Teq}
  \right]^{\frac{1}{q-1}+\frac{3(N-2)}{2}},
\end{align}
where each $p_j$ is positive for $j=3N-5, \cdots, 3N$.
Certain conditions must be satisfied for the previous calculation to be valid. 
One is the condition $N>2$. The others hold when Eq.~\eqref{eqn:cond_denom_prob:1} is satisfied.

There is functionally relationship: 
\begin{align}
  f^{[2]}(\vec{p}_{N-1}, \vec{p}_{N}) =
  \frac{1}{2^6} \left( \frac{L}{\hbar \pi} \right)^6 
  \left( \sum_{n_1, \cdots, n_{3(N-2)}=1}^{\infty} P(n_1, \cdots, n_{3N}) \right) .
\end{align}
Using the equation $U=3N\Teq/2$, we obtain
\begin{align}
  f^{[2]}(\vec{p}_{N-1}, \vec{p}_{N})
  =& \frac{1}{(2\pi m\Teq)^3}
  \frac{\left(\frac{1-q}{q}\right)^4}{\left[1-\left(\frac{1-q}{q}\right) \frac{3N}{2}\right]^4}
  \frac{\Gamma\left(\frac{1}{1-q} - \frac{3(N-2)}{2}\right)}{\Gamma\left(\frac{q}{1-q} - \frac{3N}{2}\right)}
  \nonumber \\ & \times
  \left[ 1+
  \frac{\left(\frac{1-q}{q}\right)}{1-\left(\frac{1-q}{q}\right) \frac{3N}{2}}
  \frac{(\vec{p}_{N-1})^2+(\vec{p}_{N})^2}{2m\Teq}
  \right]^{\frac{1}{q-1}+\frac{3(N-2)}{2}} ,
  \label{f2}
\end{align}
where $p_j$ for $j=3N-5, \cdots, 3N$ satisfies $-\infty < p_j < \infty$ in Eq.~\eqref{f2}.

The correlation function $\Corr(\vec{p}_{N-1}, \vec{p}_{N})$ is defined by
\begin{align}
\Corr(\vec{p}_{N-1}, \vec{p}_{N}) = \frac{f^{[2]}(\vec{p}_{N-1}, \vec{p}_{N})}{f^{[1]}(\vec{p}_{N-1}) f^{[1]}(\vec{p}_{N})} . 
\end{align}
The deviation of the corelation function from unity indicates statistical dependence.
Using Eqs.~\eqref{eqn:momdist:2} and \eqref{f2}, we obtain
\begin{align}
  \Corr(\vec{p}_{N-1}, \vec{p}_{N})
  =& \frac{\Gamma\left(\frac{q}{1-q} - \frac{3N}{2} +4\right) \Gamma\left(\frac{q}{1-q} - \frac{3N}{2} + 1\right)}
       {\left[ \Gamma\left(\frac{q}{1-q} - \frac{3N}{2} +\frac{5}{2}\right) \right]^2}
  \nonumber \\ &
  \times
  \frac{
    \left\{ 1 + \frac{1}{\left[ \left(\frac{q}{1-q}\right) -  \frac{3N}{2} \right]} 
    \frac{(\vec{p}_{N-1})^2 + (\vec{p}_{N})^2}{2m\Teq} \right\}^{\frac{1}{q-1} + \frac{3(N-2)}{2}}
  }
  {
    \left\{ 1 + \frac{1}{\left[ \left(\frac{q}{1-q}\right) -  \frac{3N}{2} \right]} 
    \frac{(\vec{p}_{N-1})^2}{2m\Teq} \right\}^{\frac{1}{q-1} + \frac{3(N-1)}{2}}
    \left\{ 1 + \frac{1}{\left[ \left(\frac{q}{1-q}\right) -  \frac{3N}{2} \right]} 
    \frac{(\vec{p}_{N})^2}{2m\Teq} \right\}^{\frac{1}{q-1} + \frac{3(N-1)}{2}}
  }.
\label{eqn:Corr}
\end{align}
For $q<1$, the correlation function is not one for free particles even in the presence of the well-known relation $U=3N\Teq/2$.
This property arises from the fact that
the statistical independence does not hold at $q \neq 1$:
$f^{[2]}(\vec{p}_{N-1}, \vec{p}_{N}) \neq f^{[1]}(\vec{p}_{N-1}) f^{[1]}(\vec{p}_{N})$.
As expected, when $\vec{p}_{N-1}$, $\vec{p}_{N}$, and $N$ are fixed, we can show the following equation: 
\begin{align}
\lim_{q\rightarrow 1^{-}} \Corr(\vec{p}_{N-1},\vec{p}_N) = 1.
\end{align}

Setting $\vec{p}_{N-1} = \vec{0}$ in Eq.~\eqref{eqn:Corr}, we have 
\begin{align}
  \Corr(\vec{0}, \vec{p}_{N})
  = \frac{\Gamma\left(\frac{q}{1-q} - \frac{3N}{2} +4\right) \Gamma\left(\frac{q}{1-q} - \frac{3N}{2} + 1\right)}
       {\left[ \Gamma\left(\frac{q}{1-q} - \frac{3N}{2} +\frac{5}{2}\right) \right]^2}
    \left\{ 1 + \frac{1}{\left[ \left(\frac{q}{1-q}\right) -  \frac{3N}{2} \right]} 
    \frac{(\vec{p}_{N})^2}{2m\Teq}
    \right\}^{-\frac{3}{2}} . 
\label{eqn:special_case}
\end{align}
The correlation function at $\vec{p}_{N-1}=\vec{0}$ goes to zero as $|\vec{p}_{N}|$ goes to infinity. 
Setting $\vec{p}_{N-1} = \vec{p}_N$ in Eq.~\eqref{eqn:Corr}, we have 
\begin{align}
  \Corr(\vec{p}_N, \vec{p}_{N})
  =& \frac{\Gamma\left(\frac{q}{1-q} - \frac{3N}{2} +4\right) \Gamma\left(\frac{q}{1-q} - \frac{3N}{2} + 1\right)}
       {\left[ \Gamma\left(\frac{q}{1-q} - \frac{3N}{2} +\frac{5}{2}\right) \right]^2}
  \frac{
    \left\{ 1 + \frac{1}{\left[ \left(\frac{q}{1-q}\right) -  \frac{3N}{2} \right]} 
    \frac{(\vec{p}_{N})^2}{m\Teq} \right\}^{\frac{1}{q-1} + \frac{3(N-2)}{2}}
  }
  {
    \left\{ 1 + \frac{1}{\left[ \left(\frac{q}{1-q}\right) -  \frac{3N}{2} \right]} 
    \frac{(\vec{p}_{N})^2}{2m\Teq} \right\}^{\frac{2}{q-1} + 3(N-1)}
  }.
\label{eqn:special_case2}
\end{align}
The correlation function at $\vec{p}_{N-1}=\vec{p}_N$ goes to infinity as $|\vec{p}_{N}|$ goes to infinity 
when the condition $1-1/(3N/2+1) < q < 1$ is satisfied.

\begin{figure}
\centering
\includegraphics[width=0.45\textwidth]{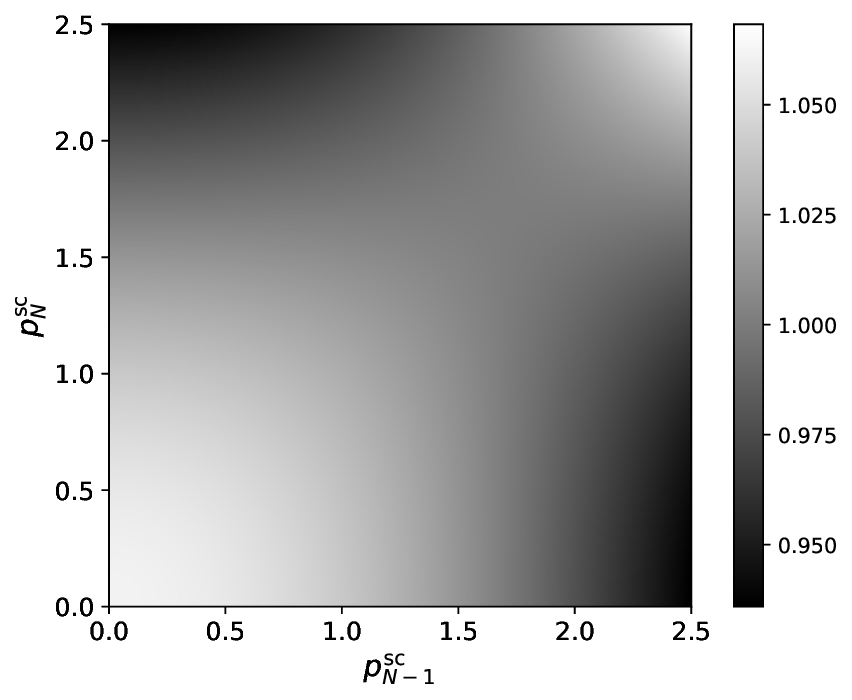}
\caption{Grayscale plot of the correlation function at $q=0.985$ and $N=20$.}
\label{fig:corr:q985:N20}
\end{figure}

For numerical illustration, we show a grayscale plot of the corelation function at $q=0.985$ and $N=20$ 
in Fig.~\ref{fig:corr:q985:N20}.
We use the scaled variables $p_{N-1}^{\mathrm{sc}}$ and $p_{N}^{\mathrm{sc}}$ which are defined by Eq.~\eqref{eqn:scaled_varible}.
The correlation function is symmetric under the interchange of $\vec{p}_{N-1}$ and $\vec{p}_N$,
and this property is clearly seen in the figure.
The correation function at $p_{N-1}^{\mathrm{sc}}=0$ decreases as  $p_{N}^{\mathrm{sc}}$ increases,
as shown in Eq.~\eqref{eqn:special_case}.
Along the diagonal line $p_{N-1}^{\mathrm{sc}} = p_{N}^{\mathrm{sc}}$, starting from the origin
$(p_{N-1}^{\mathrm{sc}}, p_{N}^{\mathrm{sc}}) = (0,0)$, the correlation function decreases, reach a minimum, and increases after that.

\section{Discussion and Conclusions}

We studied the momentum distribution and the correlation function in the $N$ free particle system
in the Tsallis statistics, which is characterized by the entropic parameter $q$, with the conventional expectation value.
The probability was obtained by applying the maximum entropy principle
and was subsequently expressed in terms of the equilibrium temperature, which is often called the physical temperature. 
The entropic parameter $q$ is less than one for the power-law-like distributions.
It should be noted that the distribution is also derived from the R\'enyi entropy. 

The relation between the energy and the temperature in the Boltzmann--Gibbs statistics holds even in the Tsallis statistics,
when the equilibrium temperature is adopted.
The relation in the canonical ensemble
is expected because the equilibrium temperature is equivalent to the temperature in the Boltzmann--Gibbs statistics in the microcanonical ensemble:
$1/\Teq = \partial \ln W/\partial U$, where the quantity $W$ is the number of states. 

In contrast, the momentum distribution and the correlation function in the Tsallis statistics for $q < 1$,
with the conventional expectation value, 
are clearly different from those in the Boltzmann--Gibbs statistics, even when the equilibrium temperature is adopted. 
The momentum distribution is power-law-like and explicitly depends on the number of particles.
The correlation function also depends on the number of particles.  
Even for free particles,
the value of the correlation function in the present statistics is different from that in the Boltzmann--Gibbs statistics. 
This property arises from the fact that the statistical independence does not hold
in the Tsallis statistics at $q \neq 1$.

The number of particles $N$ constraints the entropic parameter $q$. 
The constraints arise from the integrals. 
It is required that 
the quantity $N$ is larger than two and that the parameter $q$ satisfies the inequality $1-1/(3N/2+1)<q<1$, 
when treating the momentum distribution and the correlation function.
The deviation from the Boltzmann--Gibbs statistics $|1-q|$ goes to zero as $N$ goes to infinity. 

It is useful to compare the present results wth those obtained in Ref.~\cite{Parvan:PLA:2006, Parvan:PLA:2010} for an ideal gas.
The momentum distribution given in Ref.~\cite{Parvan:PLA:2006}, expressed in terms of the equilibrium temperature,
coincides with the momentum distribution given in the present study.
The probabilities given in Refs.~\cite{Parvan:PLA:2006, Parvan:PLA:2010} and in the present study
coincide in form when the corresponding quantities are appropriately identified. 
However,the admissible range of $q$ in the present study is narrower than that given in Ref.~\cite{Parvan:PLA:2006, Parvan:PLA:2010}.
This narrower range is related to the finiteness condition discussed in Subsec.~\ref{subsec:free-particle:probability}.
In addition, the correlation function was explicitly calculated in the present study.

We obtained the momentum distribution and the correlation function
within the framework of the Tsallis statistics with the conventional expectation value,
adopting the equilibrium temperature. 
We hope that this work will be helpful for future studies. 

\bigskip


\noindent
\textbf{Author Contributions}\quad
M.~I. performed all aspects of the study, including conceptualization, calculations, figure preparation, and manuscript writing.

\smallskip
\noindent
\textbf{Funding}\quad
This research received no specific grant from any funding agency in the public, commercial, or not-for-profit sectors.

\smallskip
\noindent
\textbf{Data Availability Statement}\quad
This study is theoretical, and all numerical results can be reproduced from the equations provided in the manuscript.

\smallskip
\noindent
\textbf{Conflict of interest}\quad
The author declares no conflict of interest.

\appendix
\section{Probability in the R\'enyi statistics}
\label{sec:appendix:Renyi}

In \ref{sec:appendix:Renyi}, the probability $\PRi$ is derived in the R\'enyi statistics.
The R\'enyi entropy is given by
\begin{align*}
  \SR = \frac{\ln \left(\sum_i (\PRi)^q \right)}{1-q}.   
\end{align*}
The functional $I_{\mathrm{R}}$ is defined by
\begin{align}
I_{\mathrm{R}} = \SR - \aR \Big[ \Big(\sum_i P_i \Big) - 1 \Big] - \bR \Big[ \Big( \sum_i P_i E_i \Big) - U \Big],  
\end{align}
where $\aR$ and $\bR$ are the Lagrange multipliers in the R\'enyi statistics. 

The following equation is obtained by requiring $\delta I_{\mathrm{R}} = 0$:
\begin{align}
\frac{q}{1-q} \frac{(\PRi)^{q-1}}{\sum_j (\PRj)^q} - \aR - \bR E_i = 0 . 
\end{align}
From this equation, we have
\begin{align}
  \aR = \frac{q}{1-q} - \bR U.  
\end{align}
For simplicity, we define $\tilde{R}$ by $(\tilde{R})^{q-1} = \sum_j (\PRj)^q$.
We obtain
\begin{align}
  \PRi = \tilde{R} \left[ 1 + \left(\frac{1-q}{q}\right) \bR \left(E_i - U \right) \right]^{\frac{1}{q-1}}.
  \label{eqn:temporary}
\end{align}
This probablilty is rewritten as follows:
\begin{subequations}
\begin{align}
  \PRi
  &= \tilde{R} \left[ 1 - \left(\frac{1-q}{q}\right) \bR U \right]^{\frac{1}{q-1}}
  \left[ 1 + \frac{\left(\frac{1-q}{q}\right) \bR E_i}{1 - \left(\frac{1-q}{q}\right) \bR U} \right]^{\frac{1}{q-1}}
  = Z_{R}^{-1} \left[ 1 - \frac{E_i}{\left(\frac{q}{q-1}\right) \bR^{-1} +  U} \right]^{\frac{1}{q-1}},  
  \label{eqn:Renyi:Parvan:Prob}
  \\
  & Z_R^{-1} = \tilde{R} \left[ 1 - \left(\frac{1-q}{q}\right)\bR U \right]^{\frac{1}{q-1}}.
\end{align}
\end{subequations}
Equation~\eqref{eqn:Renyi:Parvan:Prob} is the probability given in Ref.~\cite{Parvan:PLA:2010}.

Again, we rewrite Eq.~\eqref{eqn:temporary}.
The quaitity $\tilde{R}$ must satisfy the following equation:
\begin{align}
(\tilde{R})^{q-1} = (\tilde{R})^q \sum_j \left[ 1 + \left(\frac{1-q}{q}\right) \bR \left(E_j - U \right) \right]^{\frac{q}{q-1}}. 
\end{align}
Therefore, we have
\begin{align}
  \PRi
  = \frac{\left[ 1 + \left(\frac{1-q}{q}\right) \bR \left(E_i - U \right) \right]^{\frac{1}{q-1}}}
  {\sum_j \left[ 1 + \left(\frac{1-q}{q}\right) \bR \left(E_j - U \right) \right]^{\frac{q}{q-1}}}. 
  \label{eqn:appendix:ishihara}
\end{align}
Equation~\eqref{eqn:appendix:ishihara} is the probability given in Eq.~\eqref{p_i},
because $\Teq$ coincides with $\bR^{-1}$.

\section{Integrals}
\label{sec:appendix:Integrals}
In \ref{sec:appendix:Integrals}, we give the integrals appeared in this paper.
The following integral $K_j$ ($j = 0, 1, 2, \cdots$) appears:
\begin{align}
K_j = \int_{-\infty}^{\infty} dx_1 dx_2 \cdots dx_{3(N-j)} [1 + a(x_1^2+x_2^2+\cdots+x_{3N-1}^2 + x_{3N}^2)]^b \qquad (a>0) . 
\end{align}

Using spherical coordinates in $3(N-j)$ dimensions, the integral is evaluated.  
For $j=0$, we have 
\begin{align}
  K_0 =& \left( \frac{\pi}{a} \right)^{\frac{3N}{2}} \frac{\Gamma(-b - 3N/2)}{\Gamma(-b)} 
\end{align}
with $-b > 0$, $-b>3N/2$, and $3N>0$. 
For a positive integer $j$, we have 
\begin{align}
  K_j =& \left( \frac{\pi}{a} \right)^{\frac{3(N-j)}{2}} \frac{\Gamma(-b - 3(N-j)/2)}{\Gamma(-b)}
  \nonumber \\ & \times
  \left\{1 + a\left[ (x_{3(N-j)+1})^2+(x_{3(N-j)+2})^2+\cdots+(x_{3N-1})^2 + (x_{3N})^2\right] \right\}^{b+3(N-j)/2}
\end{align}
with $-b > 0$, $-b > 3(N-j)/2$, and $3(N-j)>0$. 


\end{document}